\documentclass[mathleft,fleqn]{an}
\usepackage{graphicx}

\begin{document}

\Pagespan{1}{}
\Yearpublication{2021}%
\Yearsubmission{2021}%
\Month{1}%
\Volume{342}%
\Issue{1}%
\DOI{asna.202100000}%

\title{The symbiotic binary ZZ CMi: intranight variability \\
and suggested outbursting nature}

\author{R. K. Zamanov\inst{1}
          \and
	  K. A. Stoyanov\inst{1}
	  \and 
	  A. Kostov\inst{1} 
          \and 
	  A. Kurtenkov\inst{1}
	  \and
	  G. Nikolov\inst{1}
	  \and
	  G. Latev\inst{1}
	  \and
          M. F. Bode\inst{2}\inst{,3} 
	  \and 
	  J. Mart\'{\i}\inst{4}
	  \and 
	  P. L. Luque-Escamilla\inst{4}
	  \and
	  N. Tomov\inst{1}
	  \and
	  Y. M. Nikolov\inst{1}	
	  \and
	  S. S. Boeva\inst{1}
}
\titlerunning{The symbiotic star ZZ~CMi}
\authorrunning{Zamanov et al.}
\institute{
        Institute of Astronomy and National Astronomical Observatory, 
        Bulgarian Academy of Sciences,  72 Tsarigradsko Shose, 1784 Sofia, Bulgaria \\
              \email{ rkz@astro.bas.bg, kstoyanov@astro.bas.bg }
         \and
               Astrophysics Research Institute, Liverpool John Moores University, IC2,
	       149 Brownlow Hill, Liverpool, L3 5RF, UK  
	  \and
	       Vice Chancellor's Office, Botswana International University of Science and Technology, 
	       Private Bag 16, Palapye, Botswana
	  \and
	        Departamento de F\'isica, Escuela Polit\'ecnica Superior de Ja\'en, Universidad de Ja\'en, 
                Campus Las Lagunillas,  A3, 23071, Ja\'en, Spain 
}

\received{11 March 2021}
\accepted{26 August 2021}
\publonline{...}

\keywords{Stars: binaries: symbiotic -- white dwarf -- Accretion, accretion discs
             -- Stars: individual: ZZ~CMi }

\abstract{%
   We present photometric and spectral observations of the symbiotic star ZZ~CMi.
   We detect intranight variability -- flickering and smooth variations in U band. 
   The amplitude of the flickering is about $0.10-0.20$ mag in U band. 
   In the B band the variability is lower, with amplitude $\le 0.03$ mag. 
   We also detect variability in the $H\alpha$ and $H\beta$ emission lines, 
   and find an indication  for outflow with velocity of about 120-150~km~s$^{-1}$.
   The results indicate that ZZ~CMi is an accretion powered symbiotic 
   containing an M4-M6 III cool component 
   with an white dwarf resembling recurrent novae and jet-ejecting symbiotic stars. 
  }
\maketitle

\section{Introduction}


The symbiotic stars are interacting binaries with long
orbital periods in the range from 100 days to more than 100 years. 
They consist of an evolved red or yellow giant transferring mass to a hot compact object
(e.g. Miko{\l}ajewska 2012). 
The mass donor is a  giant or supergiant of spectral class G-K-M.  
If the giant is an Asymptotic Giant Branch star, the system usually is a strong infrared source. 
More than 250 Galactic symbiotic systems are known (Akras et al. 2019, Merc et al. 2019).
Only a handful of them  display  flickering activity
(which is  variability on a timescale of  $\sim 10$ minutes with amplitude 
$\sim 0.2$ magnitude) -- RS~Oph, T~CrB, MWC~560, Z~And, V2116~Oph, CH~Cyg,
RT~Cru, o~Cet, V407~Cyg, V648~Car and EF~Aql. 
The last two were identified as flickering sources during the last decade: 
V648~Car (Angeloni et al. 2012) and  EF~Aql (Zamanov et al. 2017). 

ZZ~CMi  (BD+09~1633) is not a well understood object. 
Sanford (1947) classified it as an M6 star,  
noted the presence of variable $H\alpha$ emission, and 
shell type absorption in $H\alpha$ giving a velocity of -40 km~s$^{-1}$. 
Iijima (1984) noted the presence of 
high-excitation [Ne~III] and [O~III] emission lines on the objective prism spectra 
in 1982 and 1983 and classified it as a symbiotic star. 
Bopp (1984) pointed that ZZ~CMi displays significant changes in  emission lines
and  has spectroscopic and photometric characteristics similar to the symbiotic star EG~And 
 -- weak emission lines and no IR excess. 
The orbital period of ZZ~CMi seems to be about 440 days (Wiecek et al. 2010). 

Here we report optical photometry of ZZ~CMi and detection of flickering
in Johnson U band. We also report a few spectra and find variability in the emission lines. 
 
\section{Observations}
CCD photometry\footnote{The data are available upon request from the authors
and on www.astro.bas.bg/$\sim$rz/ZZCMi/ZZCMi.tar.gz} 
was obtained with the 60~cm telescope
and the 50/70~cm  Schmidt telescope of 
the Rozhen  National Astronomical Observatory (NAO), Bulgaria
and 1.23m telescope of the Calar Alto observatory, Spain. 
The journal of the CCD photometry is  given in Table~\ref{tab.obs}.
As comparison stars we used 	
TYC 764-474-1  (V 9.826, B 10.061, U 10.17)
and
TYC 763-411-1  (V 10.481, B 10.563, U 10.67).
The check stars were TYC~764-314-1 
and 
TYC 763-890-1.  
Additionally, we have 5 optical spectra\footnote{The spectra are available 
on www.astro.bas.bg/$\sim$rz/ZZCMi/ZZCMi.tar.gz}
of ZZ CMi  secured with the 
ESpeRo Echelle spectrograph (Bonev et al. 2017)
on the 2.0 m RCC  telescope of Rozhen NAO  (see Table~\ref{tab.spec}).

\begin{figure*}   
  \vspace{5.0cm}   
  \includegraphics{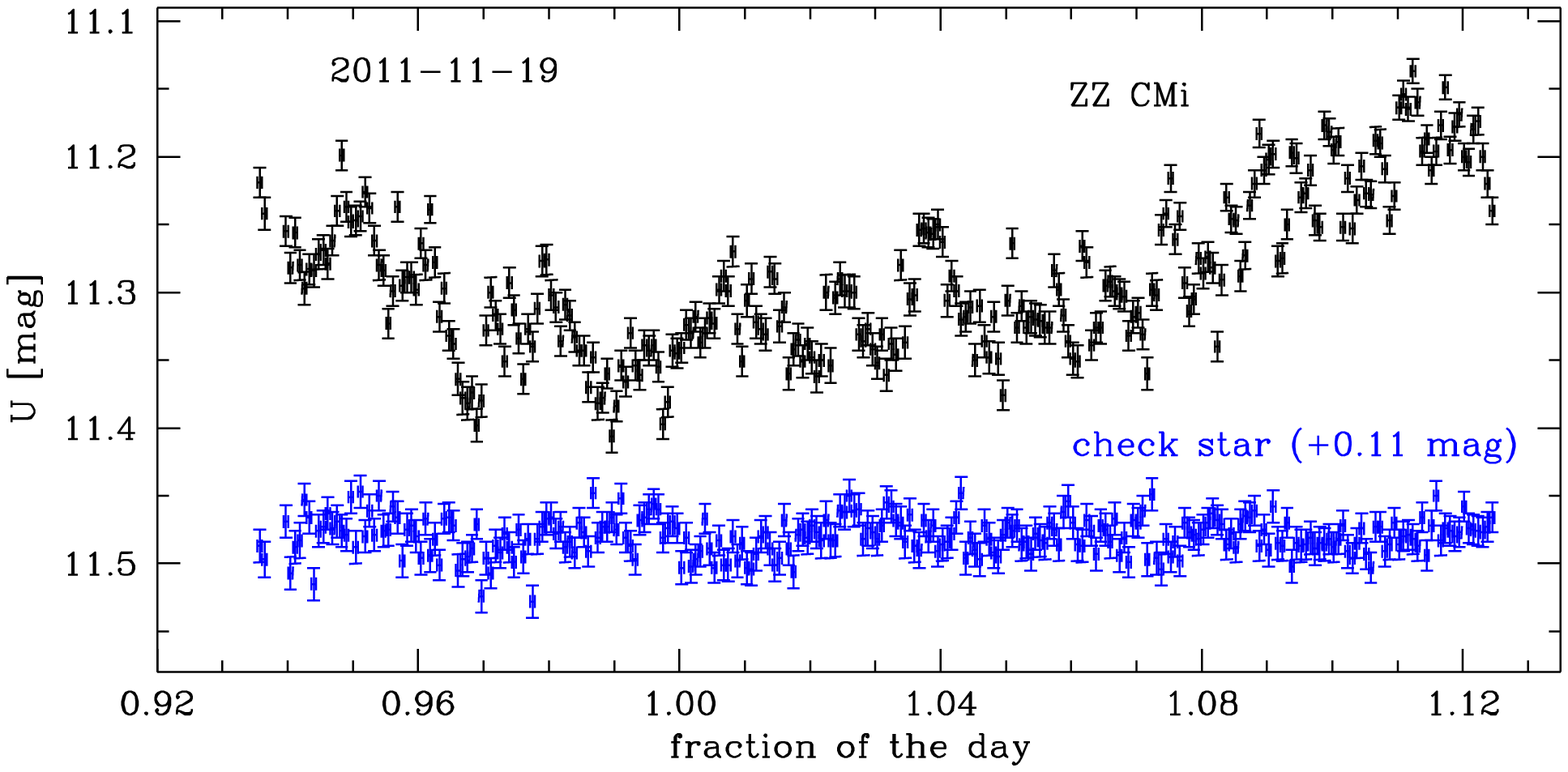}      
  \includegraphics{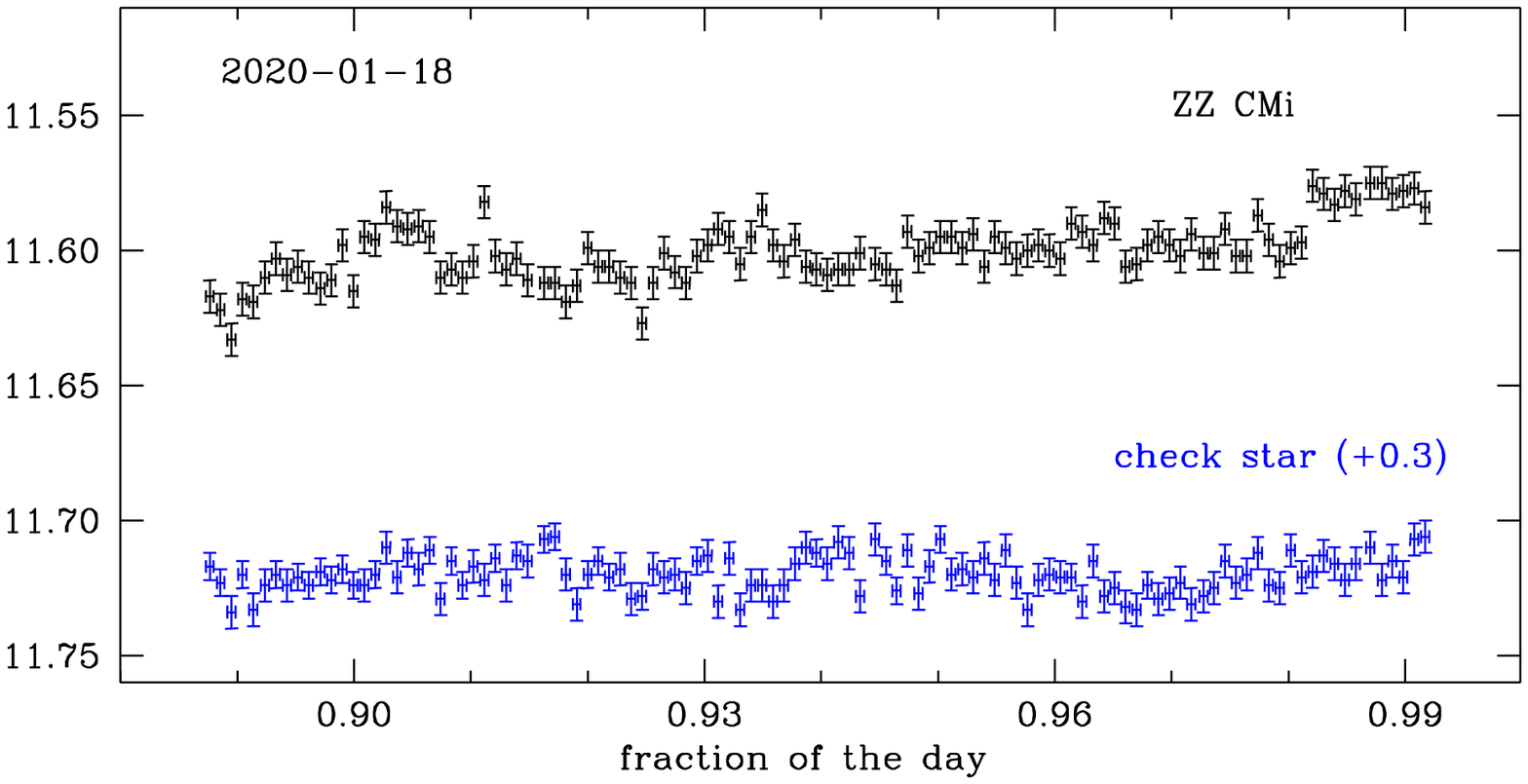}	 
  \includegraphics{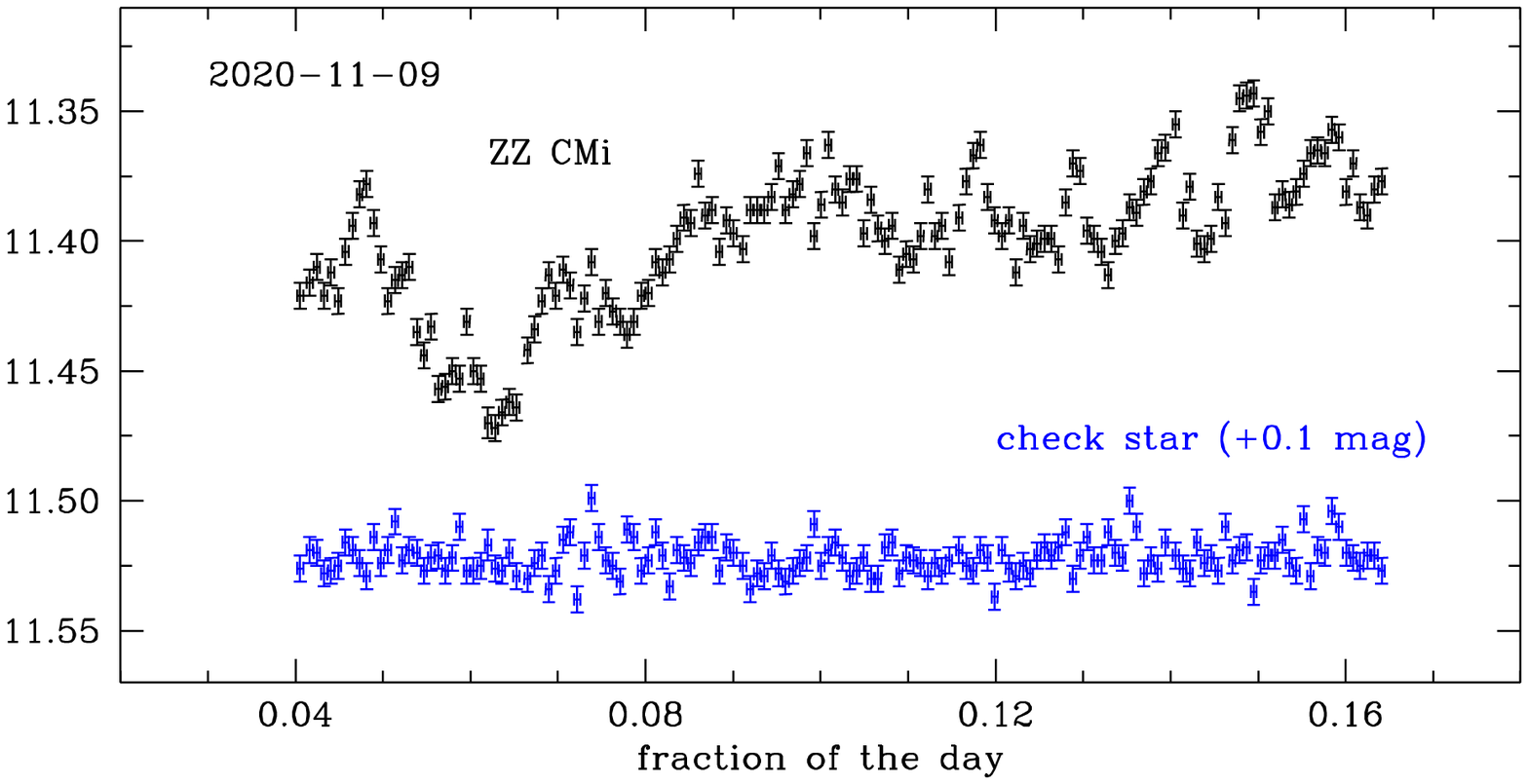}
  \caption[]{The symbiotic binary ZZ CMi -- detection of  flickering in Johnson 
  U band  on 19 November 2011 and 9 November 2020.  }
  \label{fig.1}
  \vspace{5.2cm}   
  \includegraphics{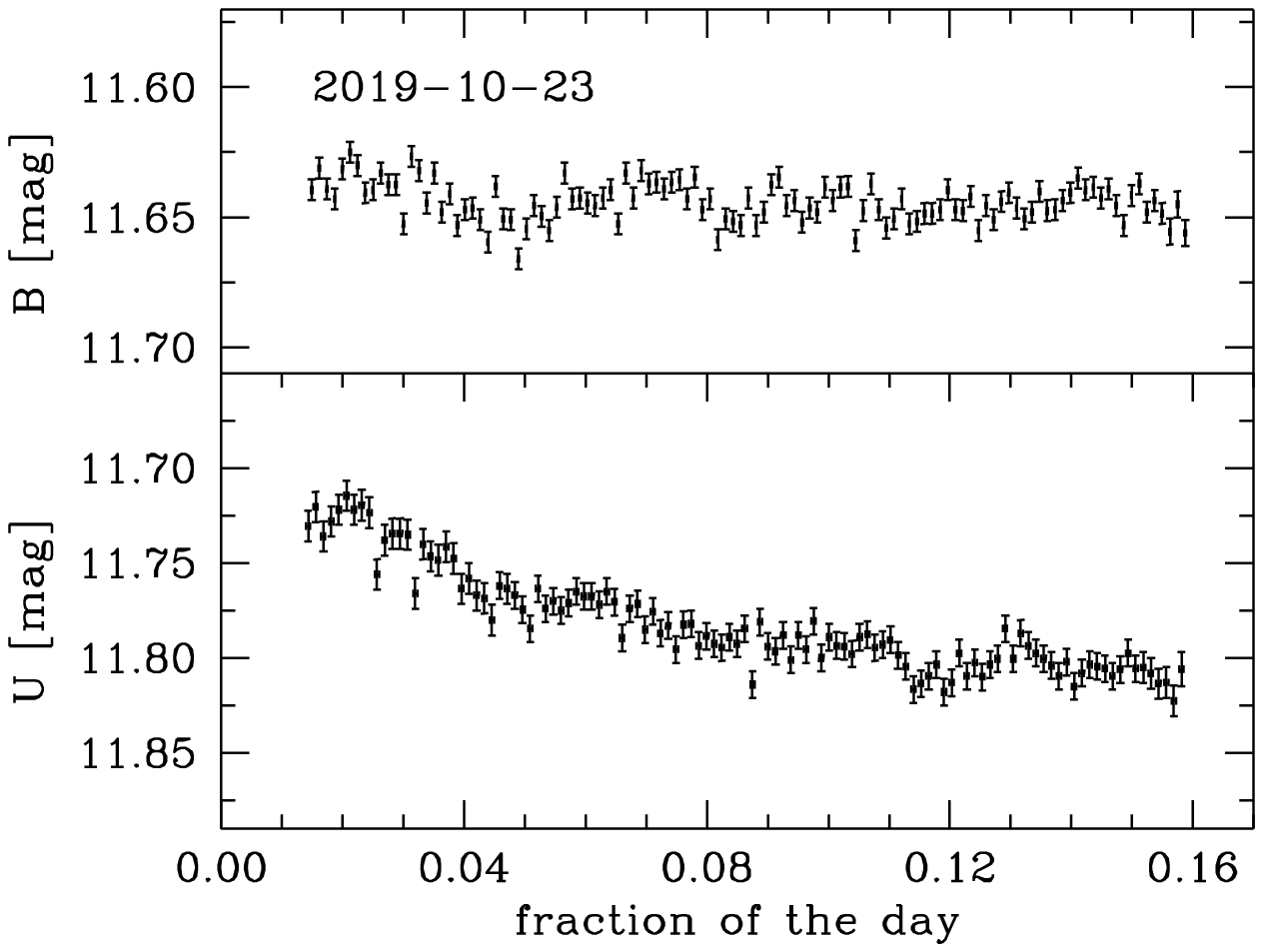}      
  \includegraphics{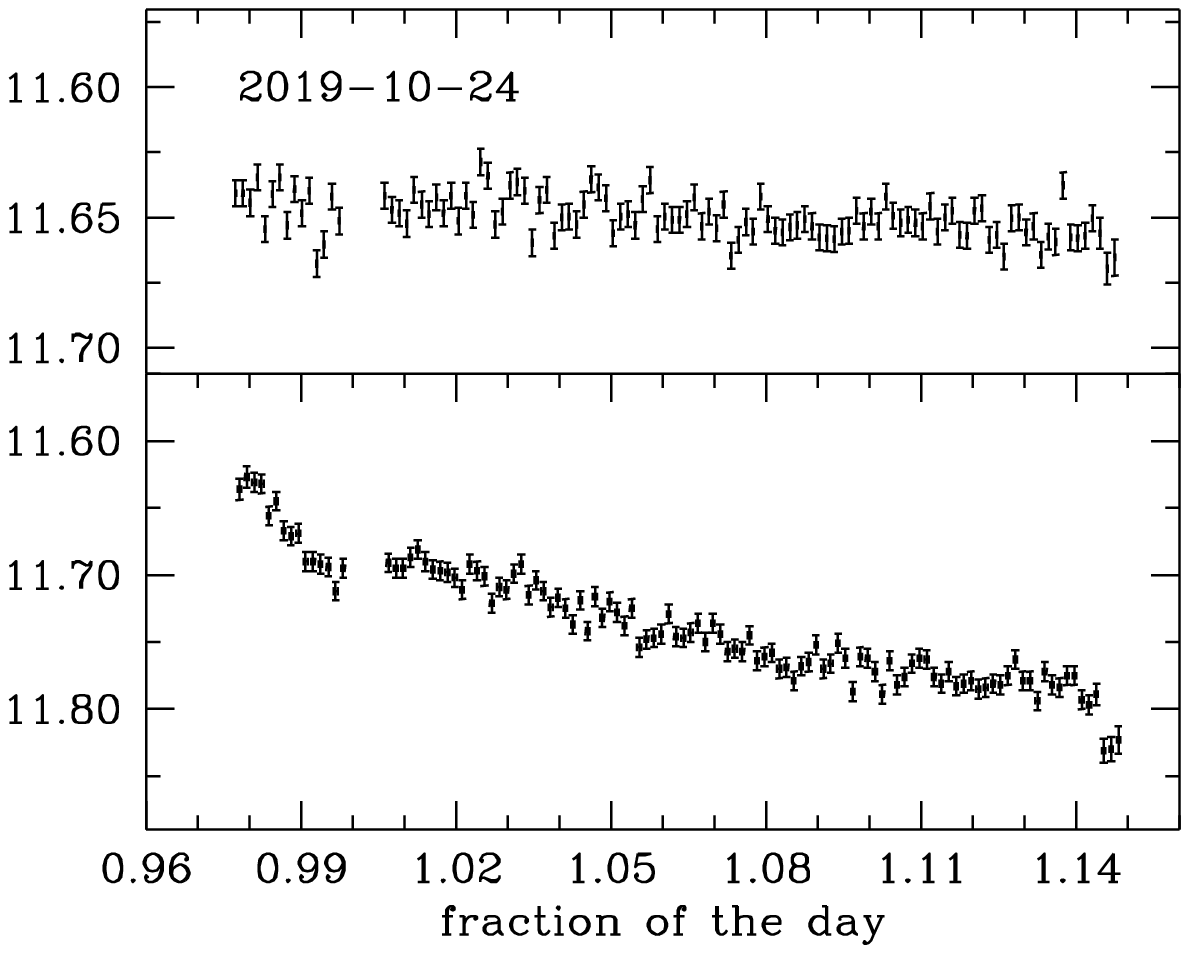}
  \includegraphics{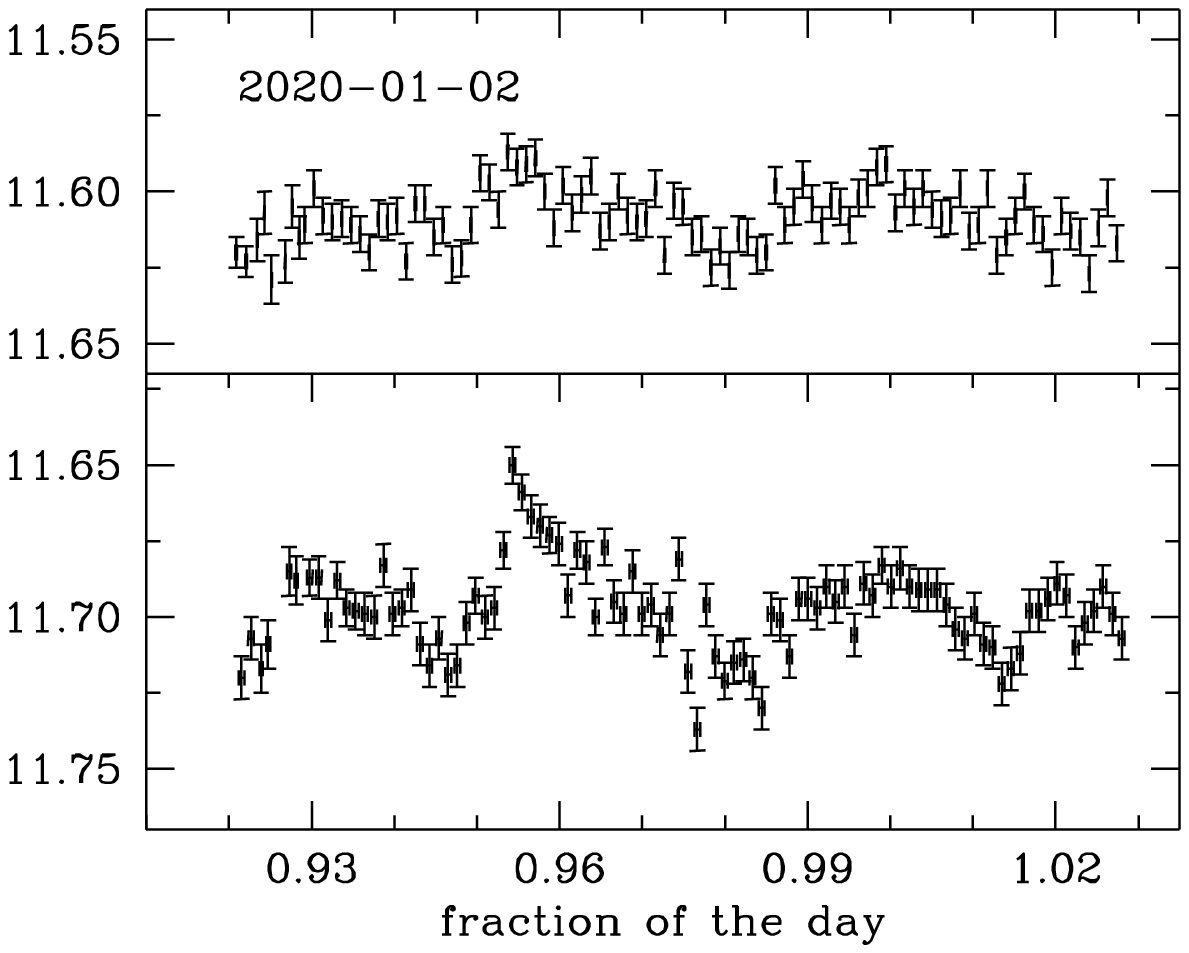}  
  \caption[]{ZZ CMi -- simultaneous  U and B band observations. 
  The intranight variability is clearly visible  only in U band.    
  }
  \label{fig.UB}
  \vspace{7.5cm}   
  \includegraphics{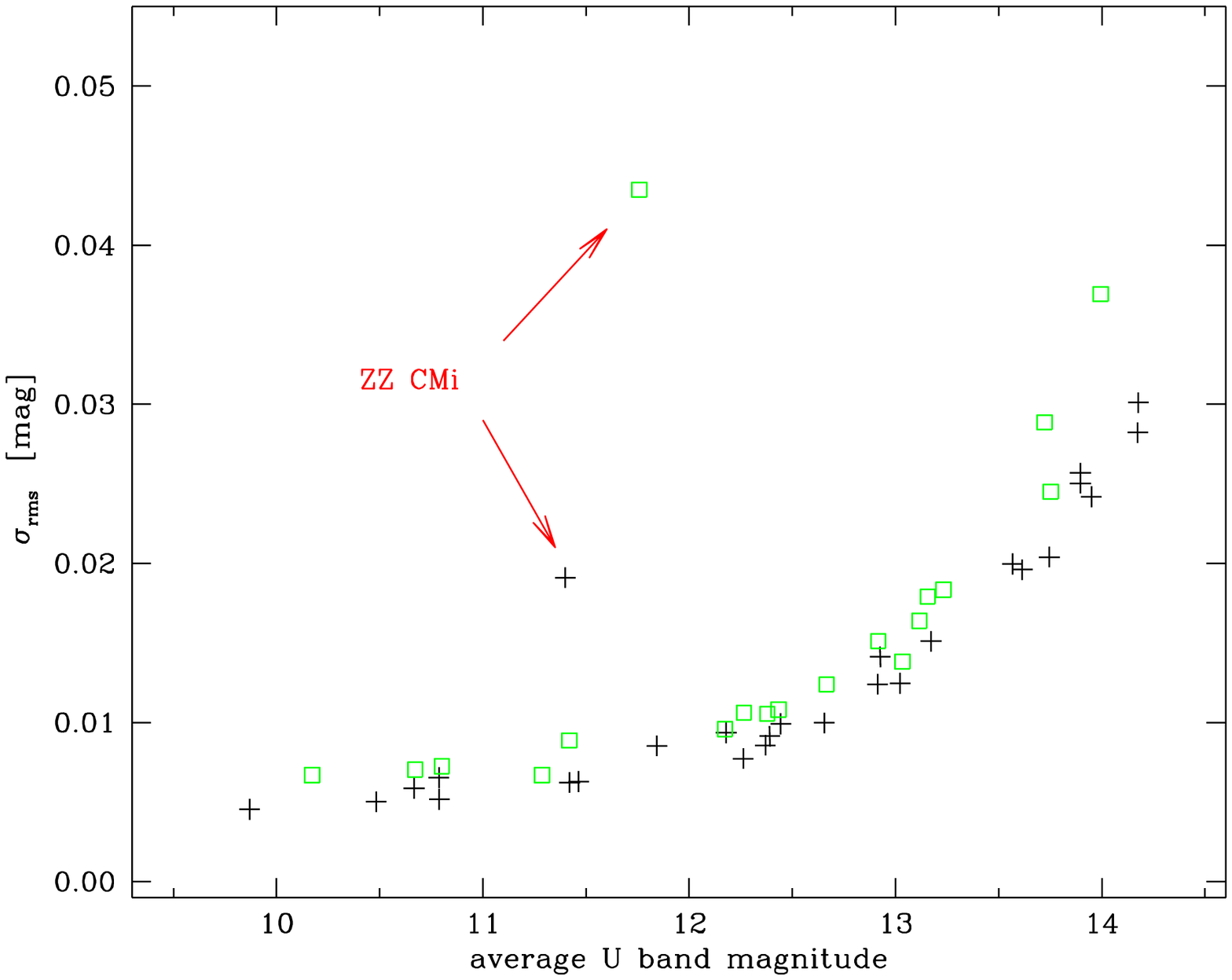} 
  \includegraphics{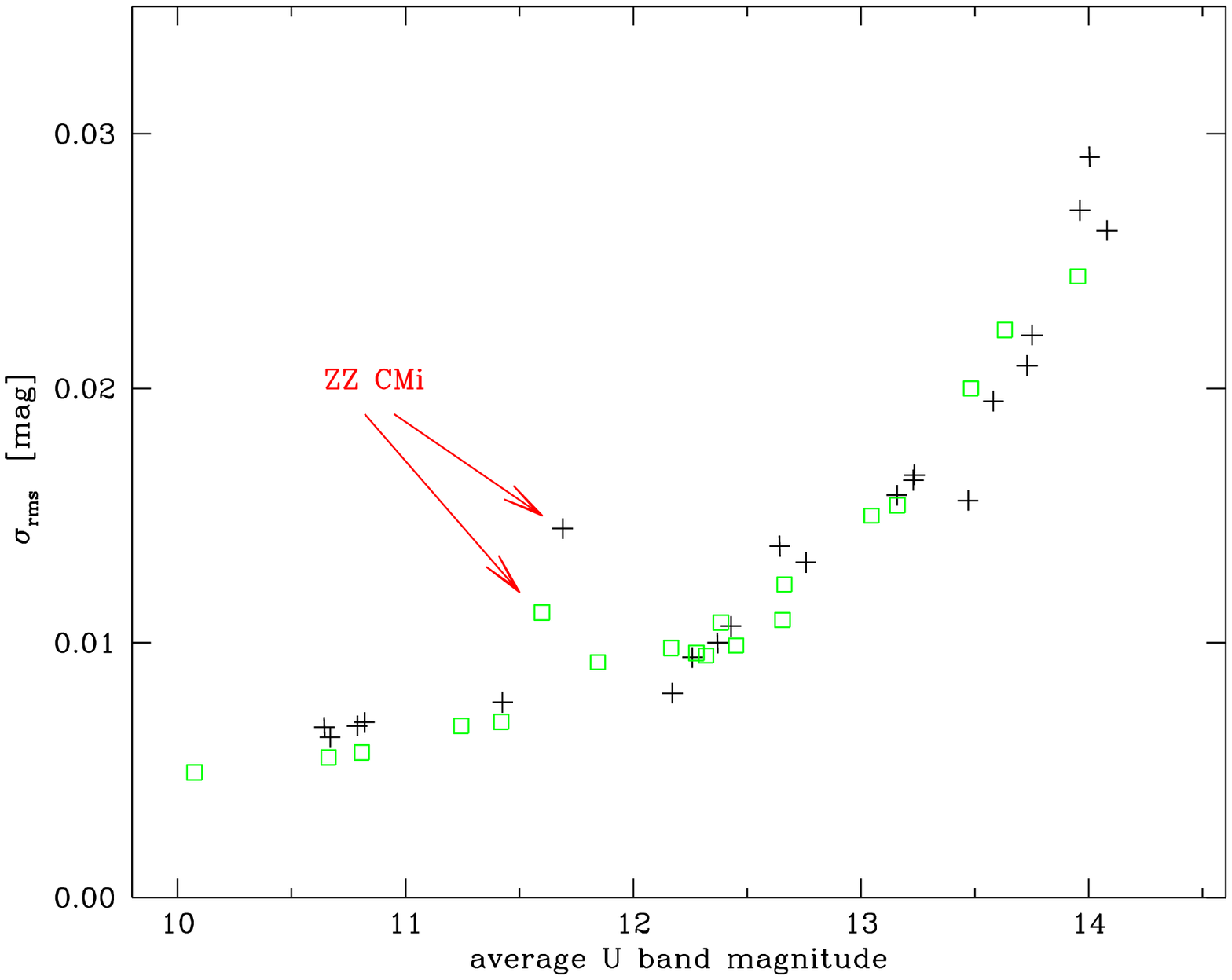} 
  \caption[]{Root mean square deviation versus the average U-band magnitude. 
             Left panel: The plus signs refer to the night 2020-11-09 and the squares to
             2019-10-24. Right panel: The plus signs refer to the night 20200102 
	     and the squares to       2020-01-18.   
	     The rms of ZZ CMi deviates from the behaviour
             of the other stars, which indicates that it is variable during our observations.
	     }
\label{fig.rms}      
\end{figure*}        

 \begin{figure}   
  \vspace{9.2cm}   
  \includegraphics{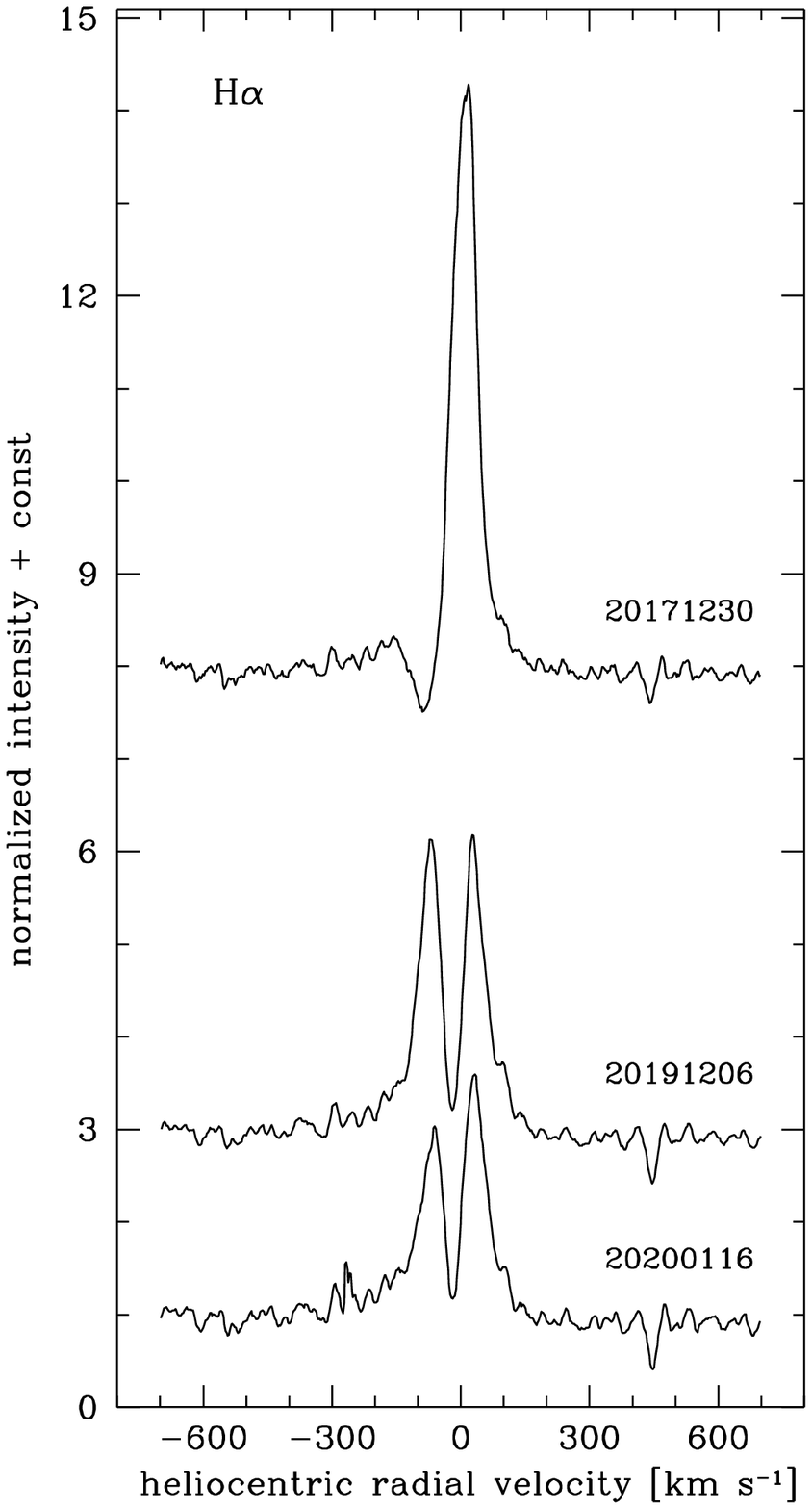} 
  \includegraphics{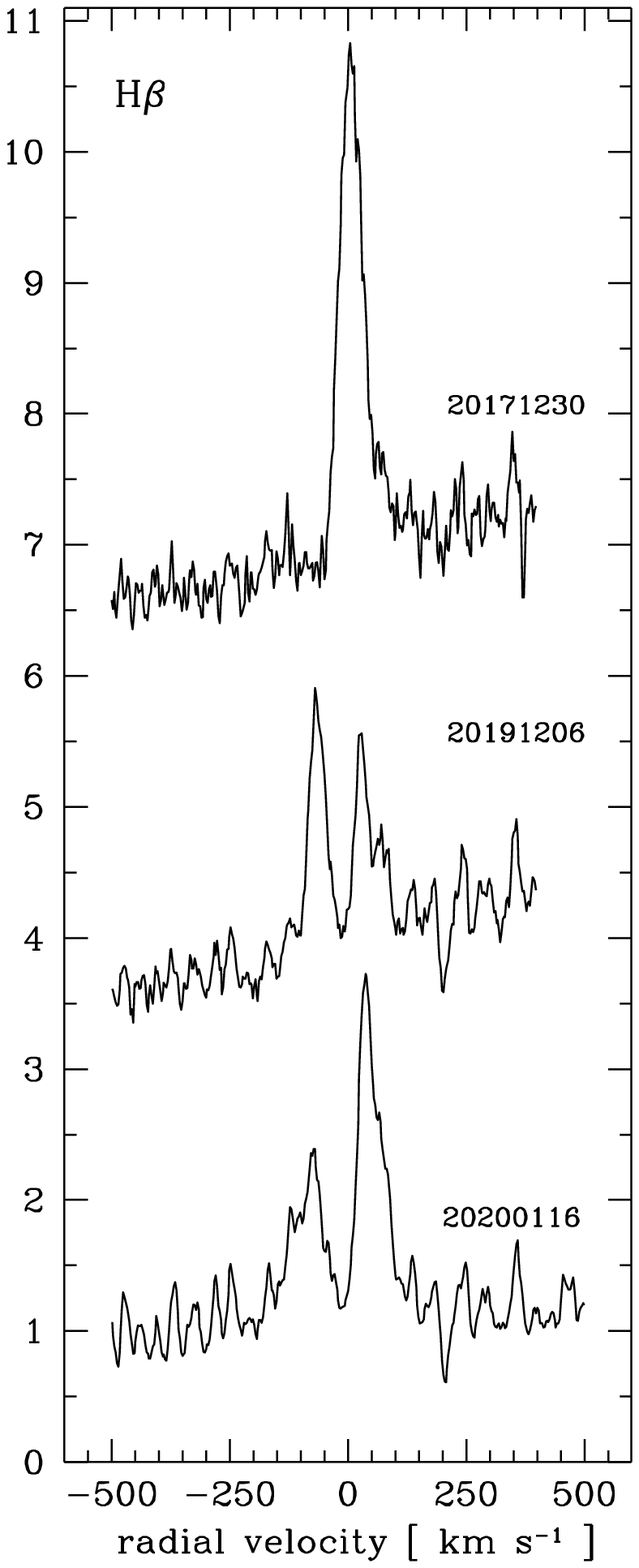} 
  \caption[]{Variability of $H\alpha$ and $H\beta$ emission lines of the symbiotic  star ZZ~CMi. 
  }
  \label{f.Ha.Hb}      
  \vspace{7.5cm}   
  \includegraphics{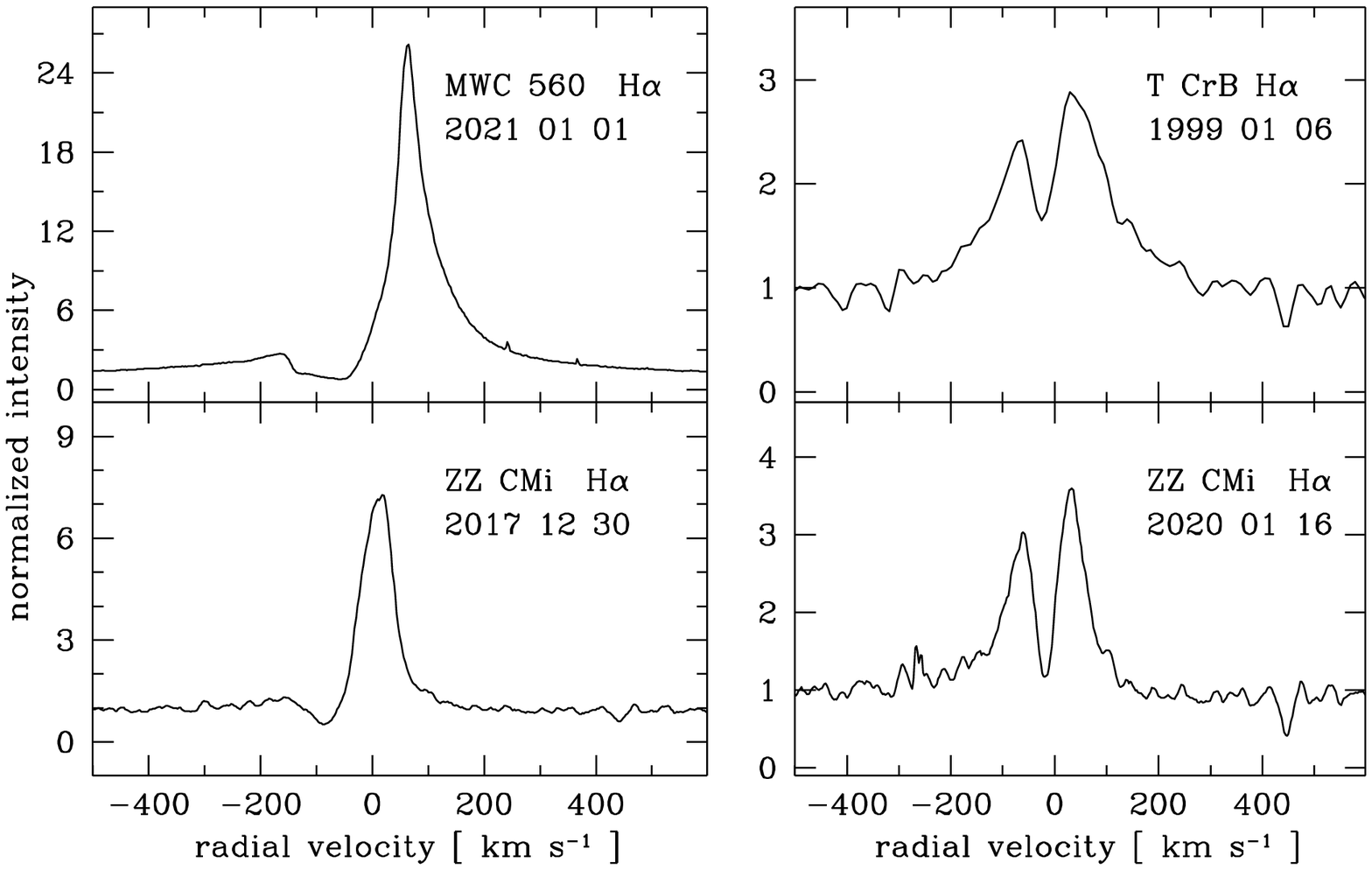} 
  \caption[]{Comparison between the $H\alpha$ emission lines of ZZ~CMi,  
  the jet-ejecting symbiotic star MWC~560 (left panel) and the
  recurrent nova TCrB (right panel). 
  }
 \label{f.MWC}      
\end{figure}         

\begin{table*}
\caption{Photometric observations of ZZ CMi. In the table are given date
(in format YYYY-MM-DD), telescope, band, UT start-end,
number of the frames and exposure time, average magnitude, typical observational error (merr),  variability and
 $\sigma_{ZZ}$ / $\sigma_{nv}$.
}             
\centering
\begin{tabular}{lclrr rcll    lllr} 
\hline
  date         &  telescope  & band   &  UT start-end  &   frames     &  $m_{av}$ & merr &     variability                 &  $\sigma_{ZZ}$ / $\sigma_{nv}$ & \\   
\hline       
& & & \\
2011-02-11     &  60cm Roz    &  B  & 23:29 - 00:33 &  71 x 30 sec & 11.52 & 0.004 &    no  			           &  0.013 /  0.011 &  \\            
               &              &  V  & 23:31 - 00:33 &  72 x 10 sec & 10.23 & 0.004 &    no  			           &  0.012 /  0.012 &  \\
2011-11-19     &  60cm Roz    &  U  & 22:26 - 02:58 & 262 x 60 sec & 11.29 & 0.011 &    yes, $\Delta U = 0.27$ mag         &  0.054 /  0.012 &  \\
2011-12-29     &  60cm Roz    &  B  & 23:40 - 00:52 & 120 x 30 sec & 11.71 & 0.010 &    no  			           &  0.025 /  0.025 &  \\  
2012-01-01     &  60cm Roz    &  U  & 22:24 - 02:08 & 217 x 60 sec & 12.00 & 0.023 &    no  (not good data) 	           &  0.030 /  0.028 &  \\	   
\\
2012-01-02     &  60cm Roz    &  V  & 23:03 - 02:36 & 400 x 30 sec & 10.55 & 0.003 &   no			           &  0.005 /  0.005 &  \\	   

2019-10-22     &  2.0m        &  V  & 23:27 - 01:23 & 277 x 3 sec  & 10.10 & 0.008 &   no                                 &  0.010 / 0.010  &  \\

2019-10-23     &  50/70cm Roz &  U  & 00:20 - 03:47 & 115 x 60 sec & 11.78 & 0.009 &   yes,  $\Delta U = 0.15$ mag         &  0.027 / 0.011  &   \\ 
               &              &  B  & 00:21 - 03:48 & 115 x 15 sec & 11.64 & 0.006 &   no			           &  0.010 / 0.009  &   \\            
& & & \\	        
2019-10-24     &  50/70cm Roz &  U  & 23:27 - 03:33 & 115 x 60 sec & 11.76 & 0.008 &   yes,  $\Delta U = 0.17$ mag         &  0.0435 / 0.0089 & \\ 
               &              &  B  & 23:27 - 03:32 & 115 x 60 sec & 11.66 & 0.007 &   no			           &  0.010  / 0.010  & \\ 
2019-12-11     &  123cm       &  B  & 00:02 - 02:31 &  62 x 10 sec & 11.42 & 0.005 &   no			           &  0.0081 / 0.0074 & \\
               &  Calar Alto  &  V  & 00:02 - 02:32 &  65 x  5 sec &  9.98 & 0.003 &   no				   &  0.026  / 0.020  & \\
	       &              &  U  & 00:06 - 02:28 &  58 x 60 sec & 12.39 & 0.009 &   yes?, $\Delta U \approx 0.04$  mag  &  0.0141 / 0.0076 & \\		
\\
2020-01-02     &  50/70cm Roz &  U  & 22:10 - 00:40 &  95 x 60 sec & 11.70 & 0.008 &   yes,  $\Delta U = 0.10$ mag         &  0.015  / 0.008    & \\
               &              &  B  & 22:10 - 00:40 &  97 x 10 sec & 11.61 & 0.006 &   yes?, $\Delta B \approx 0.02$ mag   &  0.0093 / 0.0079   & \\				   
2020-01-18     &  50/70cm Roz &  U  & 21:17 - 23:47 & 115 x 60 sec & 11.60 & 0.006 &   yes?, $\Delta U \approx 0.04$ mag   & 0.011 /  0.006     & \\  	
2020-02-02     &  50/70cm Roz &  U  & 21:15 - 22:16 & 100 x 30 sec & 11.43 & 0.015 &   no    $\Delta U < 0.04$ mag         & 0.016 / 0.016      & \\	
2020-11-09     &  50/70cm Roz &  U  & 00:58 - 03:55 & 151 x 60 sec & 11.40 & 0.005 &   yes,  $\Delta U = 0.13$ mag         &  0.0269 / 0.0070   & \\  
2021-01-20     &  60cm    Roz &  U  & 22:12 - 01:22 & 127 x 60 sec & 11.87 & 0.007 &   yes,  $\Delta U = 0.13$ mag         &  0.0304 / 0.0075   & \\  
               &              &  B  & 22:13 - 01:23 & 127 x 60 sec & 10.98 & 0.004 &   yes?, $\Delta B \approx  0.03$ mag  &  0.007 /  0.005    & \\  
 \hline                                                           
  \end{tabular}                                                  
  \label{tab.obs}
\end{table*}

\begin{table*}
\caption{Spectral observations of ZZ CMi obtained with the 2.0~m telescope of Rozhen NAO.
In the table are given date of observation, exposure time (exp-time),
parameters of $H\alpha$ line, parameters of $H\beta$ line, radial velocity of the red giant.  }             
\centering
\begin{tabular}{cc|cccc|c ccr ccc  cccccr} 
\hline
  date of observation      & exp-time & $H\alpha$ & $H\alpha$  & $H\alpha$   & $H\alpha$  & $H\beta$  & $H\beta$    & $H\beta$    &  Red Giant & \\   
yyyy-mm-dd hh:mm & [min]   &  EW [\AA] & $V_r(abs)$ & $V_r(blue)$ & $V_r(red)$ &  EW [\AA] & $V_r(blue)$ & $V_r(red)$  &	    & \\
                 &         &           &km~s$^{-1}$ &km~s$^{-1}$  & km~s$^{-1}$&           & km~s$^{-1}$ & km~s$^{-1}$ & km~s$^{-1}$& \\
2017-12-31 00:49 & 60      &   -9.1    &    -85.5   &	          &    +15.3   & -4.43     & 	         &    3.3      &  -1.3     & \\ 
2019-12-05 22:09 & 60      &   -9.2    &    -18.4   &   -67.4     &    +28.2   &  -2.59    & -69.8	 &   27.3      &   2.0     & \\
2019-12-06 21:26 & 60      &  -10.1    &    -18.7   &   -68.1	  &    +28.6   & -2.23	   & -68.7       &   25.0      &   2.1     & \\
2020-01-04 01:33 & 60      &   -8.9    &    -17.1   &   -62.1	  &    +29.9   & -5.13	   & -75.3       &   30.7      &   5.5     & \\
2020-01-16 22:20 & 60      &   -8.8    &    -18.2   &   -59.9	  &    +31.2   & -4.14	   & -77.3       &   34.4      &   5.4     & \\
 \hline                                   			
  \end{tabular}                                                  
  \label{tab.spec}
\end{table*}


\section{Results}

\subsection{Intranight photometric variability}

The results of our photometric  observations are summarized in Table~\ref{tab.obs}. 
Part of them are plotted on Fig.~\ref{fig.1} and Fig.~\ref{fig.UB}. 

For four runs we measure the standard deviation of ZZ~CMi and $\sim 15$ 
other stars in the field and plot it on Fig.~\ref{fig.rms}. 
The standard deviation is calculated as 
\begin{equation}
    \sigma_{rms} = \sqrt{ \frac{1}{N_{pts}-1} \sum\limits_{i}(m_i - \overline m )^2 } ,
    \label{eq.sig}
\end{equation}
where $\overline m$ is the average magnitude in the run, $N_{pts}$ is the number of the data points. 
$\sigma_{rms}$ calculated in this way includes 
the variability of the star (if it exists) 
and the measurement errors. 
For non-variable stars it represents  the precision of the photometry.
The standard deviation for  ZZ~CMi ($\sigma_{ZZ}$) and  standard deviation of  non-variable star  
($\sigma_{nv}$)
with similar brightness
are given in the last column of Table~\ref{tab.obs}. 
The results from  photometric observations are: 

1.  Flickering variability  is clearly 
visible in our observations in U band 
from 2011 November 19 (see Fig.~\ref{fig.1}a),  2020  November 9, 
(Fig.~\ref{fig.1}c), and 2021 January 20. 
The amplitude is 0.27 mag, 0.13 mag and 0.13 mag respectively. 

2. A decrease of the brightness in U band of 0.1 mag for 2 hours is visible in
the observations 2019 October 23  and 2019 October 24
(Fig.~\ref{fig.UB}). 
The simultaneous observations  in these two nights
do not indicate similar changes in B band. 
It is likely that the red giant is the dominating source in B band (see Sect.~\ref{red.giant}).

3. In the observations from 2020 January 18 (Fig.~\ref{fig.1}b), 
flickering is not clearly visible. 
The root-mean square (Fig.~\ref{fig.rms} right panel) 
indicates that it exists, however its amplitude
is comparable with the observational errors.

The rms deviation of ZZ CMi is 2-5 times larger
than that expected from observational errors, indicating that ZZ~CMi is variable on a timescale of 
$\sim 1$ hour. 
The presence of flickering in U band with  the observed amplitude 
strongly suggests that the hot component is a  white dwarf
(see Sect. 4.1  in  Sokoloski \& Bildsten 2010).

\subsection{Variability of the emission lines}

In Table~\ref{tab.spec} are given the following parameters measured on our spectra -- 
the equivalent width (EW) of $H\alpha$, the radial velocity of the central dip, 
the radial velocities of the blue and red peaks of $H\alpha$, the equivalent width of $H\beta$,
the radial velocities of the blue and red peaks of $H\beta$
and the radial  velocity of the red giant. 
The accuracy of the EW is about $\pm 5$\%, and that of  the velocity $\pm 1$~km~s$^{-1}$.
The EW($H\alpha$) is about  -9 \AA, which is similar 
to that of  the recurrent nova  T~CrB  in 1999 (Zamanov et al. 2005). 
Perhaps most intriguing is the profile of $H\alpha$  on 2017 December 30 which  is of P~Cyg type. 
The  absorption has centrum at -85~km ~s$^{-1}$ and extends to $-145 \pm 5$~km ~s$^{-1}$. 
This indicates an outflow with velocity of about 120-150~km~s$^{-1}$.

\subsection{System parameters}

{\bf Interstellar reddening E(B-V): }
In our spectra,  no signs of the  diffuse interstellar bands 
at 5780 \AA, 5797 \AA, and 6613 \AA\ are visible, which implies  that
the interstellar reddening to ZZ~CMi  is low.
The upper limit of the expected reddening is  $E(B-V) \le 0.04$.
This upper limit is set from our spectra as well as from the 
interstellar reddening through the  Milky Way   
calculated by IRSA: Galactic Reddening and Extinction Calculator 
in the NASA/IPAC Extragalactic Database (NED).  


{ \bf Red giant: } 
\label{red.giant}
GAIA DR2 (Bailer-Jones et al. 2018)  gives a parallax  $0.9686  \pm 0.1097  $~mas, which  
means  distance $d \approx 1030$~pc  ($927  \le d \le 1165$~pc). 
ZZ~CMi has V band magnitude in the range  $9.5 \le m_V \le 10.5$.
This gives absolute magnitude $-0.57 \le  M_V \le 0.43$. 
SIMBAD  gives  M6I-IIep  for ZZ~CMi (Shenavrin et al. 2011).
Taranova \& Shenavrin (2001) suggest that the donor star is an M4.5-5 giant. 
Straizys \& Kuriliene (1981) give for  M5 stars 
$M_V = -0.1$  for giants  and $M_V$  from $-4.7$ to $-6.7$ for supergiants.
It means that 
the classification M4-6~III is more appropriate for the cool component of ZZ~CMi.

It is worth noting that in a symbiotic binary 
a contribution to the V band from the accretion disc  and  the nebula can be expected (see e.g. Skopal 2005).
However in our observations the flickering is not visible in V band, no [OIII] $\lambda$ 5007 emission is detectable and  
the $H\alpha$ emission is weak. These suggest that the M giant is the dominating source in V band.


For an M6 giant,  
Houdashelt et al. (2000)  gives  U-V=3.234  and  B-V=1.537.  
ZZ~CMi has V band magnitude in the range  $9.5 \le m_V \le 10.5$.
This implies that the B band  magnitude of the red giant is 
$11.0  \le  m_B \le  12.0$ and  its U band magnitude is  $12.7  \le  m_U  \le  13.7$.

\subsection{ Brightness in U band}

For  ZZ~CMi  the apparent magnitude in U band is in the range 11.15 -- 12.3 mag. 
With  $d \approx 1030$~pc  and E(B-V)=0, we estimate that the absolute U band 
magnitude lies in the range $-4.2 \le M_U \le -3.0$ mag.

For comparison we can use the two symbiotic recurrent novae  T~CrB and RS~Oph, the jet-ejecting
symbiotic MWC~560 and the classical symbiotic stars AG~Dra and Z~And.
 For these objects the long term photometry is available in 
Seker{\'a}{\v{s}} et al. (2019), 
Skopal et al. (1992, 1995, 2012), 
Tomov et al. (1996) and 
Zamanov \& Zamanova (1997). 
For T~CrB the apparent magnitude in U band 
is in range $10.1 \le m_U \le  13.0$ mag.
With GAIA parallax $1.213 \pm 0.049$~mas, and E(B-V)=0.05  (Munari et al. 2016),
we estimate  $-4.7 \le M_U \le -1.8$ mag. 
For RS~Oph the apparent magnitude in U band is $10.1 \le m_U \le 13.0$ mag.
With GAIA parallax $0.442 \pm 0.053$~mas and E(B-V)=0.69 (Zamanov et al. 2018),   
we estimate $-8.8 \le M_U \le -6.6$ mag.
For MWC~560 the apparent magnitude in U band is in the range  $9.0 \le m_U \le 11.0$ mag.
With GAIA parallax $0.3534 \pm 0.1659$~mas, 
and E(B-V)=0.15 (Lucy et al. 2020)  
we estimate $M_U$ from  -9.0 to -7.0 mag.
For  AG~Dra with GAIA parallax $0.3411 \pm 0.0003$~mas, 
apparent U band in the range  $8.5 \le m_U \le 11.5$
(Leedj{\"a}rv et al. 2004) and  E(B-V) = 0.06  (Mikolajewska et al. 1995)
we estimate   $-9.2 \le M_U \le -6.2$ mag.				
For Z And with GAIA parallax $0.4866 \pm 0.0005$, 
apparent U band in the range $8 \le m_U \le  11$
and E(B-V)=0.33 (Parimucha \&  Va{\v{n}}ko 2006)
we estimate $-10.2 \le M_U \le -7.2$ mag. 

This indicates that the luminosity of the
hot component of ZZ~CMi in U band is similar to that of T~CrB. 
									


\section{Discussion}

Rapid variability is a powerful tool to 
study the  hot  companions to cool red giants and  asymptotic giant branch stars. 
The nature of the companion to Mira 
-- the prototypical pulsating giant  -- 
has been a matter of debate for more than 20 years
(e.g. Reimers \& Cassatella 1985;  Kastner \& Soker 2004).
The analysis of the rapid optical brightness variations in B band  
provided evidences that Mira B is a white dwarf (Sokoloski \& Bildsten 2010).  
The observations of Y Gem show strong flickering in the UV continuum on timescales 
of  20s, characteristic of an active accretion disc  (Sahai  et al. 2018).  
Rapid brightness variations can also be used to diagnose  
the state of the accretion disc - e.g. CH~Cyg (Sokoloski \& Kenyon 2003). 

The first indication that ZZ~CMi has intranight variability 
is given by 
Stoyanov (2012).   
These data and the new data presented  here indicate that in ZZ~CMi 
the flickering is difficult to detect in B band, because the amplitude of 
the intranight variability in B band is $\le 0.04$ mag.  
The intranight variability is visible in U band but sometimes with low amplitude. 
The searches for flickering
by Dobrzycka et al. (1996) 
  and  
   Sokoloski et al.  (2001) 
were performed in B band. Our results for ZZ~CMi
indicate that 
short time scale light variations in 
symbiotic binary stars have more chance to be detected in U and u' bands. 
The smooth variation  of the brightness 
of ZZ~CMi on  2019 October  23 and 24 (Fig.~\ref{fig.UB})
might be similar to the light curve of CH Cyg
obtained on 1997 June 9 [see Fig. 1 of (Sokoloski \& Kenyon 2003)], 
during which time the inner disc was probably disrupted. 

Depending on the main source of the energy the symbiotics can be divided in two groups: 
(1) Symbiotics with steady nuclear burning, in which the mass accretion rate is high enough 
to maintain (fuel) steady nuclear burning;  
and (2) Accretion powered symbiotics, in which 
the mass accretion rate is below the limit of steady nuclear burning and
the energy source is the accretion. 
The accretion powered symbiotics can display 
recurrent nova outbursts -- the best examples probably are  RS Oph and T~CrB (Anupama 2013) 
and collimated jets -- e.g. CH~Cyg,  MWC~560, PN Sa 3-22 (Leedjarv 2004).  

Flickering variability  is a phenomenon typical for 
the cataclysmic variables [e.g. Bruch  (1992) and references therein]
 and is also detected in
the accretion powered symbiotics.   
The only case when rapid variability 
is observed in a symbiotic with steady nuclear burning is the 
22 min periodicity in Z~And (Sokoloski \& Bildsten 1999).
This is probably due to the rotational period of 
a magnetic white dwarf, which somehow modulates the 
energy output of the nuclear burning.  

Our expectations were that all the accretion powered symbiotics should 
display flickering when the accretion disc is active. 
In accordance with these expectations, the flickering of the recurrent nova RS~Oph
disappeared after the nova outburst (Zamanov  et al. 2006)
and reappeared  240 days after it (Worters et al. 2007).
The disappearance was in the low state,
when the accretion disc was destroyed by the nova outburst
and the reappearance is a result of the resumption of the accretion. 
However the disappearance of the optical flickering of CH~Cyg 
(Sokoloski et al. 2010; Stoyanov et al. 2018)  
and 
of MWC~560 
(Goranskij et al. 2018; Zamanov et al.2020)
in a bright state as well as the behaviour of ZZ~CMi
indicate that there  
are different  mechanisms that suppress the appearance of rapid optical fluctuations. 
These mechanisms might be 
(1) spherically symmetric accretion without the formation of an accretion disc
or
(2) the parts of disc where the flickering is 
generated are in a stable state, without fluctuations, 
although another mechanism might be at work

In Fig.~\ref{f.MWC} (left panels) we plot the $H\alpha$ emission line
of ZZ~CMi and the jet-ejecting symbiotic  MWC~560. The  $H\alpha$ emission line of ZZ~CMi 
observed on  2017  December 30 resembles that of MWC~560 
obtained recently. MWC~560 is a spectacular symbiotic having a jet along the line of sight 
(Tomov et al. 1990).
The other spectra of ZZ~CMi display double-peaked profiles of  $H\alpha$ emission.
This resembles those of T~CrB in 1999 (Fig.~\ref{f.MWC}, right panels)
and can be considered 
as coming from the accretion disc around the white dwarf. 
Another possible interpretation could be similar to that on Fig.~3 by Tomov et al. (2013)
and Fig.~4 by Ikeda \& Tamura (2004), where fitting with a few gaussian components
is applied.

ZZ~CMi is also an X-ray active symbiotic from the $\beta$/$\delta$-type  (Luna et al. 2013),
which means that there are two thermal  X-ray components -- soft and  hard. 
The jet ejecting symbiotic stars 
CH~Cyg and MWC~560 which also display optical flickering are of the same  X-ray type.

Inspection of the position of ZZ CMi in the NRAO  VLA Sky Survey (Condon et al. 1998) 
reveals no radio
source detection with a 3 sigma upper limit of 1.6 mJy at 20 cm. 
The 150 MHz radio sky  survey with the Giant Metrewave Radio Telescope (Intema et al. 2017) 
also does  not detect ZZ CMi  with a 3 upper limit of 8 mJy. 
Future radio mapping with more sensitive interferometers would thus be desirable. 

With the similarities between T~CrB and ZZ CMi noted above, it may be worthwhile to
look for long term variability of the latter system in the photographic archives.

\section{Conclusions}

We report detection of intranight brightness variations from the symbiotic star ZZ~CMi.
In three nights -- 19 November 2011, 9 November 2020  
and 20 January 2021 --
flickering with amplitude $\ge 0.12$~mag in U band
is observed.  This is typical for accreting white dwarfs.   
In two other nights a smooth trend (0.1 mag for 2 hours) in brightness 
is visible. 
The spectra indicate that an outflow with velocity $\sim 120$~km~s$^{-1}$
is present sometimes. 
In our opinion ZZ~CMi is an interesting object similar to outbusting sources 
that deserves more attention from  observers. 

\section{Acknowledgements}
This work is part of  the project K$\Pi$-06-H28/2 08.12.2018  
"Binary stars with compact object"  (Bulgarian National Science Fund).
It is based on observations collected at Rozhen  National Astronomical Observatory (Bulgaria) 
and  at Centro Astron\'omico Hispano 
en Andaluc\'ia (CAHA) at Calar Alto, operated jointly by 
Instituto de Astrof{\'i}sica de Andaluc\'ia (CSIC) and 
Junta de Andaluc\'ia. JM and PLE are supported 
by grant PID2019-105510GB-C32 / AEI / 10.13039/501100011033 
from the Agencia Estatal de Investigaci\'on of the Spanish Ministerio de Ciencia, 
Innovaci\'on y Universidades entitled 
{\it High energy sources with outflows at different scales: observation of galactic sources}. 
They also acknowledge support by Consejer\'{\i}a de Econom\'{\i}a, Innovaci\'on, Ciencia 
y Empleo of Junta de Andaluc\'{\i}a as research group FQM- 322, as well as FEDER funds.

\end{document}